\documentclass{article}

\usepackage{microtype}
\usepackage{graphicx}
\usepackage{subfigure}
\usepackage{booktabs} 
\usepackage{amssymb}
\usepackage{algorithm}
\usepackage{algpseudocode}
\usepackage{amsmath}
\usepackage{tcolorbox}
\usepackage{xfrac}
\usepackage{textcomp}

\usepackage{hyperref}

\usepackage[accepted]{mlsys2025}
\makeatletter
\renewcommand{\Notice@String}{Preprint. Work in progress.}
\makeatother

\mlsystitlerunning{Local Zero-Shot Latency Prediction for Low-End GPUs}

\usepackage{tikz}
\usetikzlibrary{arrows.meta, positioning, calc}
\usepackage{graphicx}

\begin{document}

\twocolumn[
\mlsystitle{Local Zero-Shot Latency Prediction for Low-End GPUs}

\mlsyssetsymbol{equal}{*}
\begin{mlsysauthorlist}
  \mlsysauthor{Arno Geimer}{ind}
  \mlsysauthor{Beltran Fiz}{unilu}
  \mlsysauthor{Dimitris Stripelis}{flwr}
\end{mlsysauthorlist}

\mlsysaffiliation{ind}
  {Independent Researcher, Luxembourg}
\mlsysaffiliation{unilu}
  {University of Luxembourg}
\mlsysaffiliation{flwr}
  {Flower Labs}

\mlsyscorrespondingauthor
  {Arno Geimer}{research.arge@pm.me}

\mlsyskeywords{Machine Learning, MLSys}

\vskip 0.3in

\begin{abstract}
Quantifying execution latency is an important consideration across Machine Learning. Existing latency-prediction approaches typically train ML models on traces collected from workload execution on target hardware. Collection of representative traces across many consumer GPUs is expensive both in organizing the devices and in manpower, causing existing work to mainly focus on a limited set of high-end GPUs available in cloud computing. To offset this cost, we show that a local host GPU can be configured to reproduce the execution performance of less powerful target GPUs. Through selectively modifying device parameters, we obtain locally measured latencies that closely approximate those observed on the target hardware. We reformulate the latency-prediction task: instead of fitting a prediction model to a dataset of execution traces, we identify local configurations that best reproduce target devices' performance, enabling zero-shot prediction. We evaluate our approach against real measurements from 60 computer-vision models executed on 21 consumer GPU models. Our method achieves a mean absolute percentage error (MAPE) of 14.21\%, compared with more than 25\% for comparable baselines, reducing relative prediction error by 43.6\%. Predicted and measured latencies exhibit an average Spearman correlation of 0.961 across training tasks, demonstrating that the method also accurately preserves performance rankings across devices.

\end{abstract}
]

\printAffiliationsAndNotice{}
\section{Introduction}

Knowledge of model execution latency is important in several aspects of Machine Learning. In general, it is important information when selecting hardware, choosing models, and allocating compute under cost and performance constraints~\cite{lee2025forecasting, geoffrey2021Habitat}. In decentralized ML, latency estimates are integral to the orchestration of training across heterogeneous participants~\cite{lai2021oort}. Workload allocation and coordination require execution latencies in highly heterogeneous environments. For example, participant selection and straggler mitigation depend on how long clients need to complete their assigned training task. In general Machine Learning deployment, consumer GPUs which may span multiple hardware generations and performance tiers, increasingly play a central role. Estimating how a workload will perform on such heterogeneous device sets is important for both deployment and realistic system evaluation.

Despite these developments, broad coverage of consumer GPU latency stays limited in existing benchmarks and prediction works. Direct profiling requires access to each device, making measurement expensive and difficult to maintain as workloads evolve. Learned predictors can generalize beyond their training data, but they are mostly based on older benchmark datasets, with limited exposure to consumer GPUs. Additionally, many need initial measurements on the new target devices. Therefore, latency estimation for new workloads, as well as new devices, is an expensive task both in money and time.

To infer the latencies of a large set of target devices, our approach assumes access to a single device. We hypothesize that by locally reproducing the target device, local execution eliminates the need to model complex interactions in the workload operations, while approximating the performance of the target devices nonetheless. As different devices have different properties which may influence latency, we assume that reproducing these on a local device can replicate workload behavior on the target device. We investigate whether an available GPU can estimate target-device latency by executing the workload in a controlled environment. Measurements across constraint configurations reveal low-error regions in which local training-batch latency closely matches that of a target GPU. We utilize a specification-based mapping that selects one configuration per host-target pair and reuses it across workloads. Prediction requires neither target-device measurements nor a trained predictor, making the method zero-shot with respect to the target device. We find that latencies generated this way surpass existing model-training-based approaches in accuracy on our dataset. 

A key advantage of this approach to latency prediction is that it operates directly on executable workloads. Given a training step and target device information, the workload is executed directly on the restricted host GPU, and its measured runtime becomes the predicted
target latency. Configuration selection is independent of workload specifics such as model architecture, batch size, or numerical precision, removing the need for workload-specific prediction models or target-device profiling.


Our main contributions are as follows:

\begin{enumerate}
    \item We formally define how timing workloads locally under device restrictions can be used for latency prediction on unknown target devices. We empirically demonstrate that restricted host execution provides a viable alternative to model-based method in estimating training-batch latencies on unknown target devices. 

    \item We propose a zero-shot target matching method: It does not require access to target devices, neither for target profiling nor for establishing a training dataset. We show that the method enables accurate latency estimation on a wide range of unseen target devices on a single host, reducing relative prediction error by 43.6\%.
    
    \item We evaluate the approach across model families,  precisions, batch sizes and GPU generations. We comprehensively characterize the accuracy, reliability, cross-host portability, and limitations of the approach. We show that our method outperforms established baselines, and give a detailed analysis of resulting errors and host portability.
    
\end{enumerate}


\section{Related Work}\label{sec:related}

Device latency estimates serve a wide range of applications, from comparing training times across a small set of GPUs in support of hardware selection, to modeling the performance of large, heterogeneous device populations in decentralized learning deployments.

\begin{figure}[h]
  \centering
  \includegraphics[width=\linewidth]{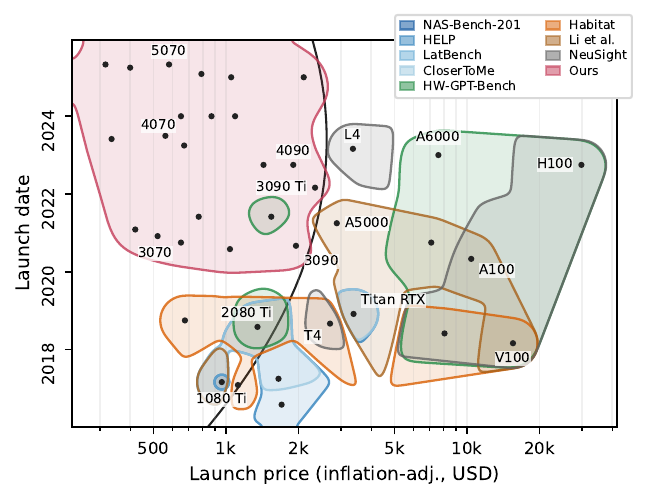}
  \caption{Evaluation GPU coverage by launch date and inflation-adjusted launch price. Shaded regions group devices by study; the black curve approximates flagship consumer GPU launch prices. This curve also shows approximately where devices become available in cloud computing services. ``Li et al.'' refers to~\cite{li2023path}. Our method and evaluation cover a large range of previously unexplored devices, focusing on cheaper and more modern targets than related works.}
  \label{fig:gpu_distribution}
\end{figure}

\paragraph{Consumer GPU coverage.}
Existing studies focus on latency prediction for server-level devices, and offer limited coverage of consumer GPUs. This is presumably due to lack of data and the absence of consumer devices in common compute rental services. Figure~\ref{fig:gpu_distribution} shows GPU coverage of existing latency predictors and benchmarks. Most evaluated devices are server-class GPUs, while represented consumer devices are mainly from older generations. Our method bridges this gap by making latency estimation possible across a broad set of consumer GPUs widely used by ML practitioners~\cite{huggingface_hardware}.

\paragraph{Latency benchmarks and predictors.}
Latency prediction aims to estimate execution times without directly measuring each target configuration. Benchmarks provide an empirical foundation for developing and evaluating such predictors. For example, NAS-Bench 101/201~\cite{ying2019bench,dong2020bench} standardize architecture evaluation, while FBNet~\cite{wu2019fbnet} incorporates hardware-aware evaluation into architecture search. LatBench~\cite{dudziak2020brp} and HW-NAS-Bench~\cite{li2021hw} provide hardware-performance data for architectures within these search spaces. Latency predictors such as HELP~\cite{lee2021hardware} and CloserToMe~\cite{tang2026closertome} build on measurements within such search spaces. However, while these NAS-oriented benchmarks provide extensive coverage within their respective architecture spaces, their coverage of workload families and GPU generations is limited.

To address this, transfer-based prediction methods, such as HELP~\cite{lee2021hardware}, LitePred~\cite{feng2024litepred}, and NASFLAT~\cite{akhauri2024latency}, adapt predictors to unseen targets using a limited set of target measurements. NeuSight~\cite{lee2025forecasting} and \textsc{Habitat}~\cite{geoffrey2021Habitat}, predict performance on unknown devices without requiring measurements from those targets. However, even NeuSight still relies on preprocessed measurements to train its predictor.

\paragraph{Simulation and resource-restricted execution.}
GPU simulators explicitly model execution and microarchitectural behaviour, including memory hierarchies and scheduling~\cite{bakhoda2009analyzing,khairy2020accel}. They support detailed architectural analysis and can be used for performance estimation, but their execution overhead can make full-workload simulation expensive~\cite{lew2019analyzing}. 

Resource-restricted native execution offers a different trade-off: the workload executes on a physical host device, while performance is modified using selected resource limits, without explicitly reconstructing the target microarchitecture. Prior work demonstrates several uses of such controls. SortingHat~\cite{namkoong2025sortinghat} downclocks GPUs to create a heterogeneous testbed and identifies Multi-Process Service (MPS) as an alternative for approximating reduced compute capacity. GSLICE~\cite{dhakal2020gslice} uses MPS to manage inference  performance, while MISO~\cite{li2022miso} predicts MIG-partition performance from measurements under MPS restrictions. Frequency-scaling studies characterize how GPU clock speeds affect model execution~\cite{tang2019impact}. More directly,~\cite{geimer2026bouquetfl} apply resource restrictions for heterogeneous federated client emulation, and subsequent work models consumer GPU capability tiers through clock and power restrictions~\cite{tajwar2026splats}. None of these works measures how accurately the restricted host reproduces target latencies.

The use of resource restrictions for GPU sharing and heterogeneous testbed construction has briefly been touched in previous works. However, effectiveness of the method on a broad set of devices and models has not been established. We close this gap by formalizing the method and demonstrating its efficacy in latency prediction on general image classification workloads across consumer GPUs.

\section{Method}

Latency prediction on unknown devices remains challenging: transfer learning-based predictors require existing reference datasets and are largely evaluated on older or server-class hardware. GPU simulators, meanwhile, are mainly designed for architectural analysis rather than execution latencies. Resource-controlled native execution appears to offer a promising alternative, but its use for latency prediction remains largely unexplored. We motivate and formalize the approach, describe the resulting optimization problem, and evaluate a basic solution across a broad range of modern consumer GPUs.
 
\subsection{Local restrictions for target reproduction.}

To motivate the design of our approach, we detail the hardware characteristics that influence execution latency, broadly distinguishing between memory-related and arithmetic performance characteristics.

\paragraph{Memory:} GPU memory affects both workload feasibility and execution speed. VRAM must hold the model parameters themselves, batched inputs, gradients, and intermediate quantities. Big models, or large inputs, may lead to out-of-memory (OOM) errors, resulting in failure to train the model. Memory bandwidth, as well as caches, also influence latency. Although VRAM frequency can be adjusted through \texttt{nvidia-smi}, our tested host devices offer only coarse settings, and preliminary experiments showed no improvement in latency estimation. We therefore leave it unchanged, memory bandwidth and cache are unmodeled in our current approach.

\paragraph{Arithmetic.}
Both its number of streaming multiprocessors (SMs) and the frequency of their execution units influence the arithmetic performance of a GPU~\cite{ernst2021performance}. Upon submission, kernels are split into thread blocks which are then scheduled to be executed in parallel across available SMs. Therefore, execution latency does not necessarily scale linearly with peak arithmetic throughput: small kernels might not need to use available SMs, while a larger workload could be forced to execute thread blocks in succession once parallel capacity is exhausted. We focus on approximating the target device's arithmetic performance, as it can be controlled through multiple hardware settings. This is supported by previous work, which has identified arithmetic throughput as a major performance bottleneck for many neural network layers~\cite{lym2019delta}. In learned latency predictors, both hardware characteristics have been used as features, further supporting their choice as representatives of device performance~\cite{tang2026closertome,lee2025forecasting}.

\paragraph{Restricting host execution.}
To effectively reduce host arithmetic performance, we use two functionalities of \textsc{Nvidia} devices. \textsc{Nvidia} MPS limits the fraction of GPU compute resources provided to a CUDA process~\cite{jain2019fractional,bakita2025hardware}. Although it does not physically disable SMs, it constrains the parallel execution capacity available to the workload. To constrict core speed, we use the \texttt{nvidia-smi --lock-gpu-clocks} command to restrict cores to a maximum clock. 

\begin{figure}[h]
  \centering
    \includegraphics[width=0.9\linewidth]{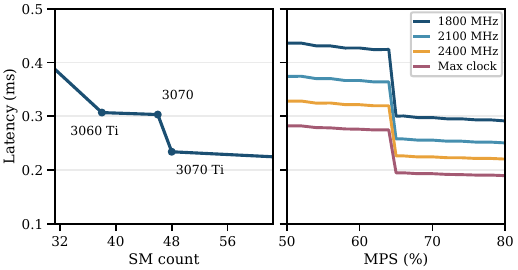}
    \caption{Latency of a convolution gradient calculation on real RTX 30-series GPUs (left) and locally under MPS and core-clock restrictions (right). MPS produces a similar step-like latency, while core-clock changes mainly shift the curves vertically, showing the different but complementary effects of both. [4070]}
    \label{fig:sm_showcase}
\end{figure}

Figure~\ref{fig:sm_showcase} illustrates how the two features influence the latency of a large convolution gradient calculation. Across GPUs of the same generation, we see a step-like latency change with respect to SM count. We are able to simulate this response through MPS restriction on a local host device. Additionally, changing the host core frequency shifts the magnitude of measured latency while preserving the shape of the response. This effect is not a new discovery~\cite{majumdar2015taxonomy}, but motivates the choice of both features for host restriction. By controlling both characteristics separately, we intend to capture workload-dependent latency behavior that cannot be adequately represented by a single peak-throughput scaling factor. The choice of a combination of both features is further established through a preliminary ablation on 15 models and 10 target devices: We observed that restricting either just core frequency or just MPS yields worse results than a combination of both by a factor of approximately 2 (25.5 resp. 20.9 vs. 11.0\%).

Having established the motivation behind the method, we formally integrate the approach into latency prediction, and propose a baseline feature selection method for target device reproduction.

\subsection{Reformulating latency prediction}\label{sec:reformulating}

We rewrite the question asked by latency prediction under the aspect of restricted host execution. Rather than \emph{How can previously measured latencies be used to fit a predictor?}, the main question becomes \emph{How can the host GPU reproduce a target GPU's performance?} The unknown switches from learned predictor parameters to host settings. Subsequently, the prediction comes from timing a workload execution on the host rather than latency inference from a fitted model. 

\paragraph{Formal definition.}Let \(w\) be the workload and \(d\) the target GPU. Here, a workload refers to a training step characterized by parameters such as the model architecture, batch size, input shape, numerical precision, and optimizer, among others. We seek to predict the latency \(t\) of running \(w\) on \(d\); the two problem formulations below show how each approach obtains the approximated latency \(\hat t\).

\begin{tcolorbox}[colback=gray!10,colframe=black!75,colbacktitle=gray!28,coltitle=black,
 fonttitle=\bfseries,boxrule=0.5pt,arc=0pt,title={Learned prediction.}]
Given latency measurements \(D=\{(w_i,d_i,t_i)\}\), fit predictor parameters \(\theta\), then infer the latency:
\[
\begin{aligned}
\theta^* & =\arg\min_\theta\sum_{(w_i,d_i,t_i)\in D}\ell\!\left(f_\theta(w_i,d_i),t_i\right),
\\
\hat t & =f_{\theta^*}(w,d).
\end{aligned}
\]
\end{tcolorbox}

Learned latency prediction relies on measured workload latencies $t_i$ from one or more devices to train a predictor. This data-driven approach is the basis of much of the work discussed in Section~\ref{sec:related}. As workloads include various features, good generalization depends on how well these characteristics are represented in the training data. Extending prediction to significantly different workload configurations may therefore require additional profiling.

\begin{tcolorbox}[colback=blue!4,colframe=blue!45!black,colbacktitle=blue!14,coltitle=black,
 fonttitle=\bfseries,boxrule=0.5pt,arc=0pt,title={Restricted execution.}]
Given a host GPU \(h\), select host settings \(c\) that best match the target performance. Predict latency by running the workload under those settings:
\[
c^*=\arg\min_{c}\Delta\!\left(h(c),d, w\right),
\qquad
\hat t=T_h(w; c^*).
\]
\end{tcolorbox}

In contrast, restricted execution assumes access to a host device $h$ and aims to approximate target performance by controlling available execution resources. The approach is largely workload-agnostic: as long as a workload can be executed on the host under the selected restrictions, its latency can be estimated without workload-specific profiling or predictor training. All latency prediction does require is a timed execution of the workload on the host. Here, \(\Delta\) denotes an arbitrary performance difference function between host and target, and $T_h$ is a timed execution on the host. Depending on its definition, evaluating \(\Delta\) might depend on target profiling measurements, but this is not necessarily the case.

\begin{figure*}[h]
  \centering
    \includegraphics[width=0.98\textwidth]{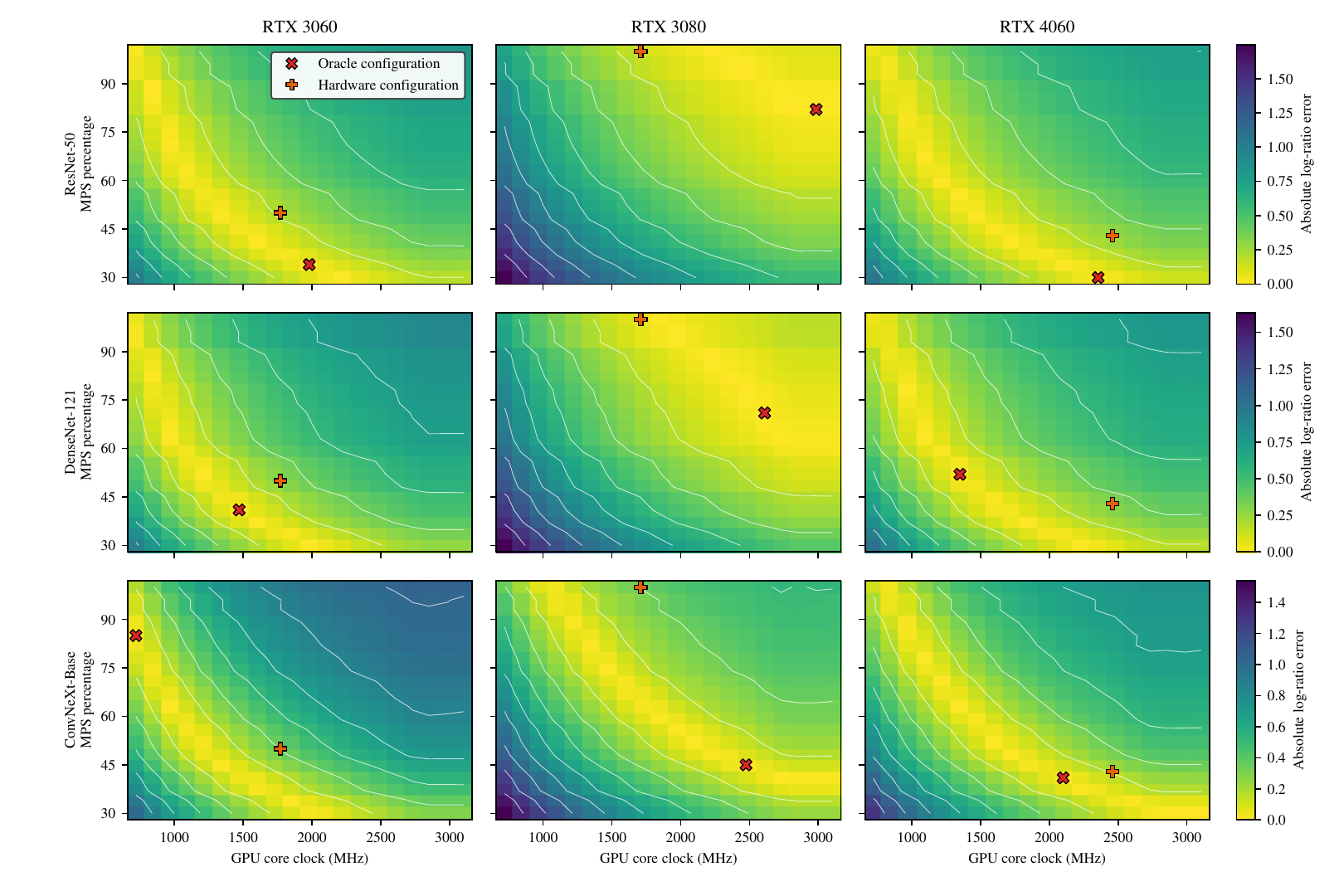}
    \caption{Training step latency error landscapes under varying MPS and GPU core-clock restrictions. Every cell compares the resulting local latency against the latency measured on the target GPU using absolute log-ratio error. Both target GPUs and models show distinct low-error valleys containing configurations that closely reproduce workload latencies. Red crosses mark the optimal minimum-error configuration in each panel, orange crosses are hardware-specification derived configurations. [4070]}
    \label{fig:valleys_workload}
\end{figure*}

\paragraph{Finding the optimum.} Having introduced the available execution controls, we next investigate whether suitable host configurations exist that closely reproduce target-device latencies. To answer this, we compare a measured target latency against predicted latency through restricted execution over the control parameters in Figure~\ref{fig:valleys_workload}. The figure shows suitable host specifications through curved regions of low error for each tested workload and target GPU. The continuous nature of these regions presumably stems from similar effective arithmetic throughput. In accordance with the observations in Figure~\ref{fig:sm_showcase}, we do observe discrete changes in error as MPS restrictions vary. Unsurprisingly, lower-performance target GPUs require more restrictive host configurations. However, the low-error regions vary across workloads, implying that a single host configuration cannot consistently reproduce target-device performance across all evaluated workloads.

We call oracle the configurations which lead to the lowest log-ratio error relative to target latency. Marked as red crosses, they show the optimal host-to-target configurations on the specific workload. Overall, on log-ratio optimized minima, oracle configurations achieve a mean absolute percentage error (MAPE) of \(0.5\%\), resulting in an exceptionally good lower bound. However, determining them assumes knowledge of the target latency. The challenge remains finding the best host settings to reproduce target performance without prior access to target measurements.

\subsection{Target mirroring estimator.}

To demonstrate the practical viability of latency prediction through restricted execution, we introduce a reproducible configuration estimator that also establishes a baseline for future work.  As previously shown, consumer target devices are difficult to access and existing profiling data are often sparse or unavailable. We therefore adopt a zero-shot approach, selecting host restrictions directly from host and target hardware specifications, without requiring target-device latency measurements.

\paragraph{Configuration mapping.} We have observed in Figure~\ref{fig:sm_showcase} that even between generations, SM count and core clock locking can reproduce kernel behaviour of target devices. We therefore employ both techniques for host-based target reproduction. We assume knowledge of both \(S_h\) and \(S_d\), the host and target SM counts, and the target's boost clock \(f_d\). 

We select the host restrictions as $\hat{p} = \frac{S_d}{S_h}$ and $
\hat{f} = \arg\min{|f_h - f_d|}$ from host-supported values, both rounded to the nearest compatible setting. Latency is then predicted through restricted workload execution: $\hat{t} = T_{h}(w; \hat{p}, \hat{f})$. The configurations selected by this method are marked as orange crosses in Figure~\ref{fig:valleys_workload}. They frequently lie outside of the low-error valleys, showcasing the limitations of a specification-based mapping. As all evaluated devices have the same number of CUDA cores per SM, the SM-count ratio is equivalent to the core-count ratios. Host memory frequency remains at its maximum setting.

\section{Results}

To evaluate the effectiveness of the approach and the accuracy of our zero-shot configuration selection, we conduct a wide evaluation focusing on four research questions. The questions address the practical requirements of latency-prediction applications.

\subsection{Experimental Setup}

\begin{figure*}[h]
  \centering
    \includegraphics[width=\textwidth]{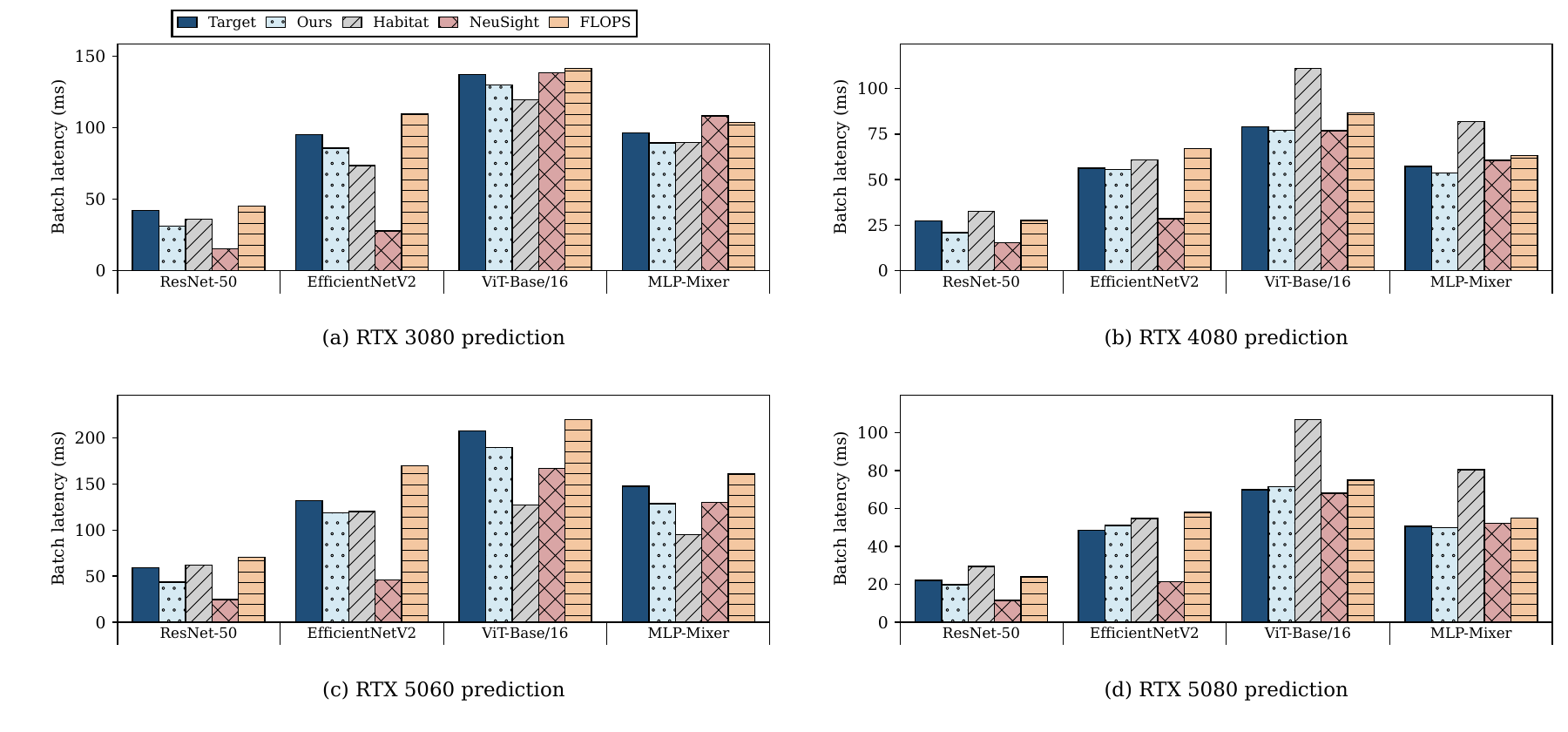}
    \caption{Batch execution latencies of different models at input size 224 $\times$ 224 across a subset of target devices. Our method closely follows target latencies, whereas baselines often over- or underestimate the target. [5090]}
    \label{fig:latencies}
\end{figure*}

\paragraph{Devices.} To build the target evaluation dataset, we rented 21 consumer GPU models through Salad~\cite{saladcloud}. Salad is not a conventional cloud compute provider: rented devices are located at independent, private hosts, not data centers. To account for varying operational environments across these hosts, we only chose devices with power limits within 95-110\% of stock. Additionally, we screened their performance against a reference dataset of averaged kernel profiles across multiple device instances, to exclude potential outliers. This ensured that devices with significant performance deviation were excluded from the dataset.

\paragraph{Host devices.}
Our primary evaluation uses an RTX 5090 rented through Vast~\cite{vastai} and evaluates the full suite of models and target GPUs. For more detailed characterization, we use a local RTX 4070 SUPER on smaller subset of 15 models and 10 target devices. Experiments using the 4070 Super mainly study individual contributions of resource restrictions, as well as the sensitivity of prediction accuracy to changes in workload parameters. In the caption of this paper's artifacts, [5090] and [4070] disclose whether results were obtained using the RTX 5090 or the RTX 4070 SUPER as hosts, or both.

\paragraph{Workloads.} We evaluate 60 image classification models from the \texttt{timm} library spanning CNNs, transformers, multi-layer perceptrons, and hybrid models combining convolution and attention. Model parameter counts range from 0.96M to 303.42M. For the main evaluation, each model is executed in FP32 with a batch size of 16 at resolutions of \(64\times64\), \(128\times128\), and \(224\times224\). This results in 180 model--resolution configurations and a total 3,780 target--workload cases, of which 3,738 are non-OOM and included in the evaluation. Out-of-memory (OOM) errors occurred only when the largest models exceeded the available VRAM capacity of lower-end GPUs. Further information on devices and workload models can be found in the Appendix.

\paragraph{Measuring latency.} To exclude data loading and host-to-device transfer from measurements, model inputs and labels are randomly generated GPU tensors. Python dependencies for both the workloads and GPU runtime are pinned. A fully timed training step consists of the forward pass, loss, backward pass, and an SGD update. After initialization and warm-up steps, we report the minimum per-step GPU time across ten timed CUDA-graph windows as workload latency to obtain optimal device performance. Target latencies are therefore optimal training-step latencies, but are not necessarily representative of full training-loop wall-clock times which include data loading and transfer. For a fixed GPU and input resolution, the slowest execution step is usually between $30.6\times$ and $83.0\times$ slower than the fastest, depending on input resolution. Code is available at \url{https://anonymous.4open.science/r/mlsys-7E84/}.

\paragraph{Error quantification} We use different measures to assess workload correctness: APE is the absolute prediction error a of singular latency error relative to the target value expressed as a percentage. Its mean over a range of workloads is the MAPE. The Spearman correlation $\rho$ evaluates prediction rankings with respect to the targets. Finally, we use the Pearson correlation measure $r$ to assess overall direction of correlation. For Pearson, we use log-transformed values to reduce bias from the larger target latencies.

\paragraph{Baselines} The approaches most similar to ours are \textsc{Habitat}~\cite{geoffrey2021Habitat} and \textsc{NeuSight}~\cite{lee2025forecasting}. While also enabling zero-shot prediction on unseen target devices, both are based on models trained on existing latency measurements. \textsc{Habitat} combines kernel profiling on the host with a wave-scaling model to predict execution latency on unknown GPUs. We adapt \textsc{Habitat} to our workloads and target devices using its published pretrained model checkpoint. \textsc{Habitat} aggregates separately timed operations, while our evaluation measures complete training steps through CUDA graphs. To account for the difference between both timing procedures, we normalize \textsc{Habitat} to the host by rescaling its predictions relative to the locally observed differences in timing measurements. This normalization improves \textsc{Habitat}'s predictions by over 80\% on our dataset.

\textsc{NeuSight} predicts model execution latency by combining kernel-level performance predictions with unseen target-device hardware specifications. We use its publicly available implementation, applying the memory-throughput-based estimation proposed in the original paper for unsupported layer types.

Because certain model architectures are incompatible with the implementations of \textsc{Habitat} and \textsc{NeuSight}, the comparison between all three approaches is restricted to 51 models supported by all baselines, rather than the full set of 60.

Additionally, we use FLOPS-based scaling as a baseline. The approach scales measured host latency by the ratio of host to target FP32 throughput, approximated using CUDA-core counts and nominal boost frequencies. 

\subsection{How well are target latencies predicted?}

We first evaluate the overall accuracy of restricted execution against target latencies and the baselines. Accurate rankings support hardware selection and workload allocation, while accurate latency prediction is essential for realistic simulation of device populations. This evaluation concerns the full target dataset of 180 workloads on 21 target devices, constructed through constrained execution on the RTX 5090.

Figure~\ref{fig:latencies} gives an overview of predicted batch latencies across different models on 224 $\times$ 224 inputs. Overall, the proposed method leads to more accurate predictions, with FLOPS scaling diverging substantially. We note the previously mentioned large discrepancy in target latency scales based on both input dimension, as well as model choice.

Table~\ref{tab:hardware-vs-scaling} summarizes prediction results. Restricted execution achieves a MAPE of 14.21\% on the workloads shared with \textsc{Habitat} and \textsc{NeuSight}, with a median of 12.01\%. Its standalone overall MAPE on all 60 workloads is similar at 14.26\%. \textsc{Habitat} trails with 25.2\% MAPE, showing limited adaptability to the choice of workloads and target devices. \textsc{NeuSight} performs poorly, with an error three times as high as ours at 45.1\%. Confirming the observations in previous works~\cite{lee2021hardware, tang2026closertome}, FLOPS-based scaling performs badly in numeric latency estimation. Our method shows the lowest error deviation, highlighting the stability of prediction errors when compared to the baselines.

Our method shows the highest correlation of all, both in ranking and in correlation direction $r$. \textsc{Habitat} and FLOPS perform relatively well, while \textsc{NeuSight} underperforms.

\begin{table}[h]
    \centering
    \normalsize
    \caption{Method comparison over all covered target devices and models. Our method largely outperforms existing predictor-based baselines while being fully zero-shot. SD denotes the standard deviation of APE, $\rho$ and $r$ are the Spearman and Pearson correlation, respectively. [5090]}
    \label{tab:hardware-vs-scaling}
    \vspace{4pt}
    \setlength{\tabcolsep}{3pt}
    \begin{tabular}{@{}lrrrrr@{}}
    \toprule
    & \multicolumn{3}{c}{APE (\%)} & \multicolumn{2}{c}{Correlation} \\
    \cmidrule(lr){2-4}\cmidrule(lr){5-6}
    Method & Mean & Median & SD & $\rho$ & $r$ \\
    \midrule
    Ours & \textbf{14.21} & \textbf{12.01} & \textbf{11.23} & \textbf{0.990} & \textbf{0.984} \\
    \textsc{Habitat} & 25.21 & 17.79 & 34.54 & 0.974 & 0.961 \\
    \textsc{NeuSight} & 45.10 & 50.96 & 27.12 & 0.873 & 0.877 \\
    FLOPS & 52.10 & 41.55 & 44.00 & 0.970 & 0.965 \\
    \bottomrule
    \end{tabular}

\par\smallskip
\end{table}

On a general level, in our use case, restricted execution outmatches training-based latency estimation. It outperforms all three baselines both in prediction accuracy as well as overall correlation to targets. This supports use of the method for heterogeneous population latency distribution estimation as well as device choice.

\subsection{How reliable are predictions?}

Device choice applications depend on the accuracy of single latency predictions. In case of significant error variance, confidence in per-device results can not be established. Therefore, we study prediction reliability through analysis of the error distribution and best- and worst performers.

\begin{figure}[h]
  \centering
    \includegraphics[width=\linewidth]{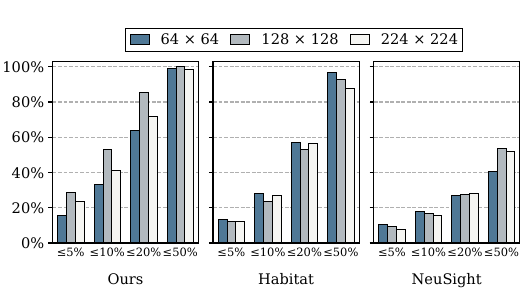}
    \caption{Share of predictions whose absolute percentage error is at or below each threshold. Our method outperforms both baselines at every threshold. [5090]}
    \label{fig:error-coverage}
\end{figure}

Figure~\ref{fig:error-coverage} shows the share of overall predictions whose APE falls below certain thresholds. Close to 80\% of restricted execution-based predictions lie within 20\% of target values, compared to less than 60\% for $\text{\textsc{Habitat}}_{hn}$ and even less for \textsc{NeuSight}. Overall, the APE 95th percentile is 34.35\%, with a 99th percentile at 47.33\% compared to Habitat and Neusight at 185.43\% and 86.8\%, respectively. The method outperforms baselines in error distribution, establishing confidence in it. 

Table~\ref{tab:performance-extrema} shows extrema: Best and worst performers across target GPUs and models. Across GPUs, prediction MAPE ranges from a minimum of 9.02\% on a target RTX 5080 up to 20.30\% on a target RTX 4060 Ti. All three top target models are the same generation as the host device, while the worst performers are from the two previous generations. Across models, MAPE ranges from 6.4\% in predicting DeIT latencies to 24.0\% in RegNet latency prediction. We note that even the worst MAPE values are lower than the average APE of \textsc{Habitat} and \textsc{NeuSight}-based predictions. Models from two architectures appear among both the best and worst performers, suggesting that model architecture seems not to play a substantial role in latency errors.

\begin{table}[h]
  \centering
  \caption{Best and worst hardware-emulation accuracy ranked by MAPE. [5090]}
  \label{tab:performance-extrema}
\vspace{4pt}
  \setlength{\tabcolsep}{4pt}
  \begin{tabular}{@{}lrlr@{}}
    \toprule
    \multicolumn{4}{c}{\textbf{Target GPUs}} \\
    \midrule
    Best & MAPE (\%) & Worst & MAPE (\%) \\
    \midrule
    5080 & 9.02 & 4060 Ti & 20.30 \\
    5060 & 11.08 & 3080 Ti & 18.38 \\
    5060 Ti & 11.69 & 4070 SUPER & 16.92 \\
    \midrule
    \multicolumn{4}{c}{\textbf{Models}} \\
    \midrule
    Best & MAPE (\%) & Worst & MAPE (\%) \\
    \midrule
    deit base & 6.44 & regnety 080 & 24.00 \\
    vit base & 7.75 & vit tiny & 23.13 \\
    twins-svt & 8.70 & deit tiny p16 & 22.84 \\
    \bottomrule
  \end{tabular}
\end{table}

Finally, a high Spearman correlation across all pairings does not imply faithful ranking for a single workload. Table~\ref{tab:device-ranking} shows aggregated Spearman correlations across workloads: For a set workload, target device ranking correlates by an average of 0.96 with the observed profiled ranking, with 90\% of workload rankings being at or above 0.913. This outperforms \textsc{Habitat}, \textsc{NeuSight}, and FLOPS. The good target ranking of FLOPS may be explained by device capability being largely correlated with latency ordering. 

While pooled Spearman correlation in Table~\ref{tab:hardware-vs-scaling} captures ranking across all workload–device pairs, per-workload metric measures device ranking for a fixed model and input resolution. It removes differences between workloads and evaluates distinctions between target devices. Note how \textsc{Habitat} and FLOPS switch places compared to Table~\ref{tab:hardware-vs-scaling}: An overall good correlation does not result in good single-device ranking performance.

\begin{table}[h]
\centering
\normalsize
\caption{Device-ranking Spearman correlations, averaged across workloads. Our method outperforms existing baselines in average and median ranking as well as ranking confidence. P10 and P90 show Spearman 10th and 90th percentiles. [5090]}
\label{tab:device-ranking}
\vspace{4pt}
\setlength{\tabcolsep}{3pt}
    \begin{tabular}{@{}lccrr@{}}
    \toprule
    Method & Mean & Median & P10 & P90 \\
    \midrule
    Ours & \textbf{0.961}{\scriptsize$\pm$0.042} & \textbf{0.977} & \textbf{0.913} & \textbf{0.989} \\
    \textsc{Habitat} & 0.843{\scriptsize$\pm$0.189} & 0.911 & 0.650 & 0.973 \\
    NeuSight & 0.866{\scriptsize$\pm$0.116} & 0.877 & 0.750 & 0.982 \\
    FLOPS & 0.960{\scriptsize$\pm$0.041} & 0.974 & 0.902 & \textbf{0.989} \\
    \bottomrule
    \end{tabular}
\par\smallskip
\end{table}

We conclude that the values obtained through restricted execution on a 5090 accurately represent target latencies, outperforming the baselines. They capture both nominal and relative latencies of target devices and workloads, and error variance is well-bounded.

\subsection{How reproducible are these results?}\label{subsec:repro}

The established results have all been computed on the same host device. However, as the method is based on local execution, different hosts may result in different latency estimates. This may pose a problem in the reproducibility of future works based on the method. Therefore, we compare the previous results to latencies generated from a host RTX 4070 Super. The reduced capability of the 4070 relative to the 5090 limits us to ten target devices instead of the previous 21: An RTX 4080 cannot be modeled reliably on a lower-end host device. 

\begin{table}[h]
  \centering
  \normalsize
  \caption{Agreement between hardware-emulation predictions produced independently on the RTX 4070 SUPER and RTX 5090. Host device choice affects predicted latencies, with predictions differing by 13\% on average. Almost 80\% of predictions differ by less than 20\%. Positive bias means the RTX 5090 prediction is slower. [4070], [5090]}
  \label{tab:host-reproducibility}
  \vspace{4pt}
\setlength{\tabcolsep}{3pt}
    \begin{tabular}{@{}llrcr@{}}
      \toprule
      Scope & \multicolumn{2}{c}{Host difference (\%)} & Pearson $r$ & $\leq$ 20\% \\
      \cmidrule(lr){2-3}
      & Mean $|\Delta|$ & Bias & & \\
      \midrule
      All & 13.19 & +3.40 & 0.990 & 79.6 \\
      64px & 15.55 & +10.20 & 0.986 & 70.7 \\
      128px & 12.05 & +5.65 & 0.989 & 85.5 \\
      224px & 11.84 & -6.59 & 0.986 & 82.9 \\
      \bottomrule
    \end{tabular}%
\end{table}

Table~\ref{tab:host-reproducibility} summarizes the latencies from both hosts and shows the differences in predictions. On average, the 4070 Super outperforms the 5090 in target latency generation with a MAPE of 11.00\% compared to 14.84\% on the ten target devices. Generated latencies differ by 3.4\% on average with an average absolute difference of 13.19\%, with slight variance depending on input resolution. Log-transformed results between both hosts are Pearson-correlated by a factor of 0.99, and almost 80\% differ by less than 20\%. The interesting observation is the bias: In lower-resolution workloads, the RTX 5090 host underestimates target performance by up to 10\% with respect to the 4070 host, while the situation switches in higher resolutions.

These results showcase an important challenge of our naive application of fixed restriction configurations per target device, irrespective of host device. Performance differences in hosts can lead to differing latency estimations, undermining the good results observed previously. However, the observed differences in host predictions are contained, and, more importantly, still show superior overall performance compared to baselines. Host-resultant differences do not invalidate the effectiveness of the method, they merely encourage further study.


\subsection{Which factors influence the validity of latency estimates?}

Finally, we conduct an ablation study on which factors influence prediction accuracy. This analysis is indispensable in developing a more adaptive restriction approach to generate more accurate predictions and handle host differences. It also helps in reinforcing the good overall results of the approach, making sure they are not just due to chosen workload specifics.

\begin{figure}[h]
  \centering
    \includegraphics[width=.75\linewidth]{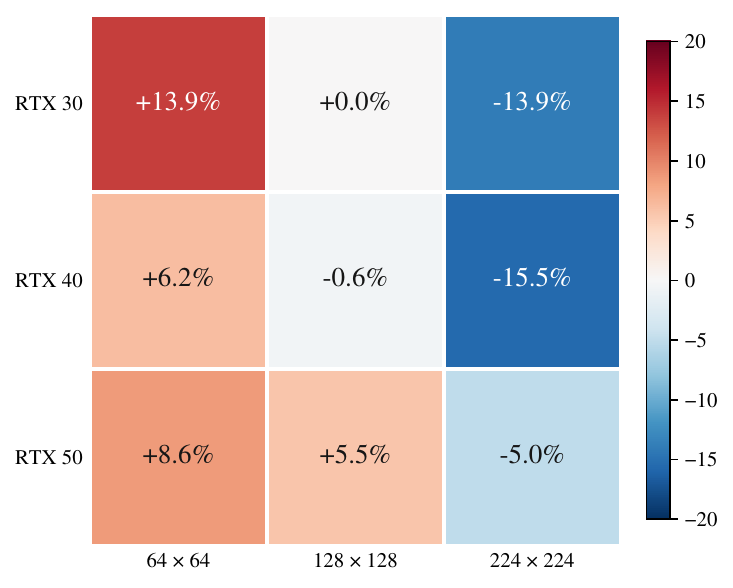}
    \caption{Prediction bias of latency estimates by target GPU generation and input resolution. Positive values mean slower execution than the target; negative values indicate faster execution. A clear difference in prediction error can be observed across both target generation and input resolution. [5090]}
    \label{fig:bias}
\end{figure}

Figure~\ref{fig:bias} shows the overall bias of predictions with respect to target devices and input resolutions. The host 5090 overestimates the latency of lower-resolution workloads, while underestimating high-resolution batch latencies. This effect was previously observed in Table~\ref{tab:host-reproducibility} relative to a better host.

Across target generations, the trend does not appear as clear: While the 5090 host shows highest bias in latency predictions against the RTX 30 generation, it performs similarly on average against both the RTX 40 and 50 generation. However, nominal bias values are less high on the 50 generation, suggesting that the host device is most consistent in approximating performance of cards from its own generation.

As the evaluation includes models from a wide range of families and sizes, Table~\ref{tab:family-mape} breaks down prediction accuracy by model architecture. Overall, MAPE values are fairly consistent, with transformers exhibiting lower prediction errors than CNNs despite having more parameters on average. While latency prediction of MLP-based models outperforms all others, the small number of MLP models tested prevents drawing a conclusion. These results suggest that neither model architecture nor size alone explain differences in prediction accuracy. 

\begin{table}[h]
\centering
\caption{Hardware-emulation MAPE by model family. Prediction errors change slightly by model family, while average model sizes are similar. Mean size is given in millions of parameters. [5090]}
\label{tab:family-mape}
\vspace{4pt}
\setlength{\tabcolsep}{4pt}
\begin{tabular}{@{}lrrr@{}}
\toprule
Family & \#Models & Mean size (M) & MAPE (\%) \\
\midrule
CNNs         & 32 & 50.03 & 15.68 \\
Transformers & 18 & 64.66 & 12.61 \\
Hybrids      &  8 & 22.89 & 14.25 \\
MLPs         &  2 & 44.38 & 10.28 \\
\bottomrule
\end{tabular}
\end{table}

\begin{figure}[h]
  \centering
    \includegraphics[width=\linewidth]{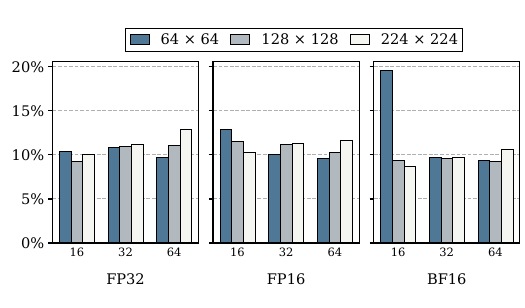}
    \caption{Characterizing MAPE resulting from different training batch size and precision format. Values are stable across both factors, with an exception at BF16 with batch size 16. [4070]}
    \label{fig:batch}
\end{figure}

Two other factors which might influence overall correctness are precision and batch size. With precision fixed to FP32 in previous experiments, both FP16 and BF16 are common in ML training. Same for the batch size, which was previously fixed at 16. Figure~\ref{fig:batch} shows MAPE values across these two factors on a smaller ablation study set. We observe limited variation in aggregate MAPE based on either variable. Changing precision or batch size does not significantly alter overall results; In fact, it improves them on average. The higher-than-usual MAPE error in BF16 precision at batch size 16 can be explained through PyTorch choosing a particularly slow kernel on the host device for two ResNet models.

Finally, across all workload--device pairings, we find negligible correlation between training-step latency and absolute percentage error (APE), with a Pearson correlation of $r=-0.0005$ on log-transformed latencies and a Spearman correlation of $\rho=-0.058$. This suggests little monotonic relationship between relative prediction accuracy and workload duration: latency alone does not explain the observed variations.

Overall, prediction accuracy remains broadly consistent across evaluated architecture families, numerical precisions, and batch sizes, although individual configurations show larger errors. Our experiments have shown that the host GPU may be biased towards its own generation.

\section{Discussion}

\subsection{Usability}

We have shown that the method can be used for a large range of applications and performs well in predicting single-GPU latencies. Combined with its workload-agnostic setup, this makes it ideal for device choice, both when the target model has already been determined, and when not.

Additionally, latency estimation of arbitrary workloads on unknown devices is useful in decentralized ML. Existing predictors are not adapted to predicting latencies for the devices employed in decentralized applications, as asserted in Section~\ref{sec:related}. Our method proposes a reasonable alternative. Our results have shown that the resultant latency distribution closely resembles a real-world population.
The assumption of access to a host device is relatively non-restrictive, as cloud GPU rental services offer viable and cheap access; for example, in our use of Vast~\cite{vastai}, an RTX~5090 could be rented for under \$1 per hour. 

The method overhead is minimal as well: host restrictions are applied in a few seconds at most, and workload execution time depends on the model and input size. In our experiments, 95\% of full profilings finish in less than 15 seconds, with very few taking more than 30 seconds, so most single latency predictions cost less than a few \textcent.

\subsection{Limitations}

Section~\ref{subsec:repro} shows the biggest weakness of the method: depending on the host device, resulting latencies can differ. We presume that this is due to our choice of configuration estimator: A fixed value, only tied to target core count and core clock, does not take low-level performance differences between hosts into account. While relatively contained, it can nonetheless lead to unsatisfactory consequences in single device choice, especially if choices are inter-generational. The bias introduced across generations and workloads, as observed in Figure~\ref{fig:bias} can misrepresent overall latencies. More refined methods, taking into account both host and target capabilities, might be able to translate more cleanly between devices.

\subsection{Future Work} We believe that differences across hosts, but also the different biases depending on host and target generation, can be improved by a cross-generational translation model for a more guided restriction choice. The current work serves as the foundation for target performance reproduction based latency prediction. The remarkably good overall results, even when employing a naive estimator of reproduction restrictions, shows the potential of the method. In future work, we plan on refining restriction parameter choice through more refined methods, such as target profiling. Formally, our choice of the $\Delta$ function in Section~\ref{sec:reformulating} is suboptimal, and we expect a more flexible performance difference estimator to lead to more accurate latency predictions.

\section{Conclusion}

In this work, we formalized latency prediction through restricted host execution. We proposed a simple zero-shot predictor for the host device, and have shown the effectiveness of the method across a large set of latency prediction tasks. The method outperforms existing works on a wide range of target \textsc{Nvidia} devices as well as image classification workloads. Our results suggest that the proposed latency prediction method is able to challenge existing data-based latency predictors.

\bibliographystyle{plainnat}
\bibliography{references.bib}

\end{document}